\documentclass[%
 aps,
 prl,
twocolumn,
superscriptaddress,
frontmatterverbose, 
showpacs,preprintnumbers,
 nofootinbib,
 amsmath,
 amssymb,
floatfix,
latexsym,array,enumerate,letter,
]{revtex4-1}
\usepackage[utf8]{inputenc}
\usepackage[english]{babel}
\usepackage{multirow}
\allowdisplaybreaks
\usepackage{graphicx,color}
\usepackage{hyperref}
\usepackage{comment}
\usepackage{appendix}
\usepackage{caption}
\usepackage{subcaption}
\usepackage{footmisc}
\usepackage{bm}
\usepackage{dcolumn}
\usepackage{gensymb}
 \usepackage{float} 
\restylefloat{table}
\usepackage{multirow}
 \usepackage{hyperref}
 \usepackage{array}

\hypersetup{
    colorlinks,
    citecolor=black,
    filecolor=black,
    linkcolor=blue,
    urlcolor=blue
}

\newcolumntype{P}[1]{>{\centering\arraybackslash}p{#1}}
 \newcommand{\be}{\begin{equation}}
\newcommand{\ee}{\end{equation}}
\newcommand{\beq}{\begin{equation}}
\newcommand{\eeq}{\end{equation}}
\newcommand{\bea}{\begin{eqnarray}}
\newcommand{\eea}{\end{eqnarray}}

\def \W  {\mathcal{W}}
\def \V  {\mathcal{V}}

\begin{document}

\title{Radial and Non-Radial Oscillations of Inverted Hybrid Stars}
\author{Chen Zhang}
\email{Corresponding author: iasczhang@ust.hk}
\affiliation{The HKUST Jockey Club Institute for Advanced Study,
The Hong Kong University of Science and Technology, Hong Kong, P.R. China}
\author{Yudong Luo}
\email{yudong.luo@pku.edu.cn}
\affiliation{School of Physics, Peking University, Beijing 100871, China}
\affiliation{Kavli Institute for Astronomy and Astrophysics, Peking University, Beijing 100871, China}

\author{Hong-Bo Li}
\email{lihb2020@stu.pku.edu.cn}
\affiliation{School of Physics, Peking University, Beijing 100871, China}
\affiliation{Kavli Institute for Astronomy and Astrophysics, Peking University, Beijing 100871, China}

\author{Lijing Shao}
\email{lshao@pku.edu.cn}
\affiliation{Kavli Institute for Astronomy and Astrophysics, Peking University, Beijing 100871, China}
\affiliation{National Astronomical Observatories, Chinese Academy of Sciences, Beijing 100012, China}

\author{Renxin Xu}
\email{r.x.xu@pku.edu.cn}
\affiliation{School of Physics, Peking University, Beijing 100871, China}
\affiliation{Kavli Institute for Astronomy and Astrophysics, Peking University, Beijing 100871, China}
\begin{abstract}
We study the radial and non-radial oscillations of Cross stars (CrSs), i.e., stars with a quark matter crust and a hadronic matter core in an inverted order compared to conventional hybrid stars. We draw comparisons of their oscillation modes with those of neutron stars, quark stars, and conventional hybrid stars. 
We find that the stellar stability analysis from the fundamental mode of radial oscillations, and the $g$, $f$ modes of non-radial oscillations are quite similar to those of conventional hybrid stars. However, due to the inverted stellar structure, the first non-radial $p$ mode of CrSs behaves in an inverted way and sits in a higher frequency domain compared to that of conventional hybrid stars. These results provide a direct way to discriminate CrSs from other types of compact stars via gravitational-wave (GW) probes. Specifically, compact stars emitting $g$-mode GWs within the $0.5$--$1$ kHz range should be CrSs or conventional hybrid stars rather than neutron stars or pure quark stars, and a further GW detection of the first $p$ mode above 8 kHz or an identification of a decreasing trend of frequencies versus star masses associated with it will help identify the compact object to be a CrS rather than a conventional hybrid star.
\end{abstract}

%\today
\maketitle
\section{Introduction}
One of the key challenges of strong interaction physics is understanding QCD matter phases at non-perturbative regimes. This regime can be explored at high temperatures by the lattice QCD studies, RHIC and future NICA and FAIR experiments, whereas the observations on compact stars could probe the behaviour of high-density matter at low temperatures. Empirically it is known that hadronic matter (HM) is bound at nuclear densities to form neutron stars (NSs). However, it is theoretically possible that compact stars are described in terms of quark degrees of freedom, a matter phase called quark matter (QM).

 The Bodmer-Witten-Terazawa hypothesis~\cite{Bodmer:1971we, Witten:1984rs, Terazawa:1979hq} suggests that QM with comparable amounts of $u, \,d, \,s$ quarks, also called strange quark matter (SQM), might be the ground state of baryonic matter at the low (zero) temperature and pressure. A recent study~\cite{Holdom:2017gdc} demonstrated that $u, d$ quark matter ($ud$QM) is, in general, more stable than SQM and the ordinary nuclear matter at a sufficiently large baryon number beyond the periodic table. Such bulk absolute stability allows the possibility of up-down quark stars consisting of $ud$QM~\cite{Zhang:2019mqb,Ren:2020tll, Wang:2019jze,Xia:2020byy, Xia:2022tvx,Cao:2020zxi,Yuan:2022dxb,Wang:2021byk,Li:2022vof,Restrepo:2022wqn}, in addition to strange quark stars consisting of SQM~\cite{Bodmer:1971we, Witten:1984rs, Terazawa:1979hq,Farhi:1984qu,Alford:2007xm,Weber:2004kj,Haensel:1986qb,Alcock:1986hz, Zhou:2017pha,Weissenborn:2011qu,Xu:1999bw,Yang:2023haz}.

The absolutely stable QM that constitutes pure quark stars is physically different from that inside the conventional hybrid stars (i.e., compact stars consisting of an HM crust with a QM core)~\cite{Alford:2004pf,Alford:2013aca,Blaschke:2022egm,Contrera:2022tqh,Paschalidis:2017qmb,Nandi:2017rhy,Christian:2018jyd,Bauswein:2018bma,Montana:2018bkb,Bauswein:2019skm,Li:2021sxb,Li:2022ivt,Miao:2020yjk,Miao:2023jqe,Constantinou:2023ged,Guo:2023som,Sotani:2023zkk},  because the latter is not stable at zero pressure and originates from the deconfinement of hadronic matter at the star center. Additionally, the HM equation of state (EOS) inside compact stars is subject to larger uncertainties due to the lack of knowledge at the intermediate pressure range~\cite{Oter:2019rqp, Kurkela:2014vha}. Therefore, considering the absolutely stable QM hypothesis, QM can transit to HM inside the quark stars if the latter becomes more stable than the former in the unexplored regions. Such matter composition will form ``Cross stars" (CrSs)~\cite{Zhang:2022pse} that consist of a hadronic matter core and a QM crust, with an inverted structure compared to the conventional hybrid stars. The formation of CrSs requires the central pressure to exceed the critical value for chemical potential crossing, which can be achieved by quark star spin-down, accretion or merger. Note that such transitions (from quark matter to hadronic matter) occur at low (nearly zero) temperatures and intermediate densities, so that they do not contradict to the conventional picture of early-universe phase transitions and the standard QCD phase diagram where things are quite uncertain in corresponding regime.

Due to the large parameter space for tuning different compact star models and the large degeneracy of global properties in the current observational mass range ($>1 M_{\odot}$), the aforementioned types of compact stars can all satisfy various recent astrophysical constraints on global properties like pulsar mass measurements, crude radius inferences, and bounds on tidal deformabilities.
On the other hand, from gravitational-wave (GW) asteroseismology, we know that GWs from the oscillation modes of compact stars can reveal rich information about their global properties, internal structure, and matter EOSs ~\cite{Andersson:1997rn,Benhar:2004xg,Lau:2009bu,Chirenti:2015dda,Sagun:2020qvc, Andersson:2021qdq,Li:2023ijg,Thapa:2023grg, Zhao:2022toc, Kunjipurayil:2022zah, Zhao:2022tcw, Lin:2013nza, Wen:2019ouw, Zhang:2012jf, Chan:2014kua, Bai:2017sar}, thus serve as a promising method to discriminate different types of compact stars in this new era of GW astronomy led by LIGO/VIRGO/KAGRA collaborations~\cite{LIGOScientific:2016aoc, LIGOScientific:2017bnn, LIGOScientific:2018mvr, LIGOScientific:2017vwq, LIGOScientific:2018hze,LIGOScientific:2020aai, LIGOScientific:2020zkf, LIGOScientific:2018cki}. 

For radial oscillations, of particular interest to our study is the fundamental ($f$) mode since a positive square of such an eigenfrequency indicates that the star is stable against radial perturbations. The studies of radial oscillations are usually categorized into slow and rapid conversions depending on the phase transition timescale compared to the oscillation timescale. For slow conversions in conventional hybrid star settings, it has been found that $\partial M/ \partial P_c>0$ is insufficient to determine the radial stability~\cite{Pereira:2017rmp,Lugones:2021zsg}. 

When compact stars oscillate non-radially, various modes of GWs are radiated, including the gravity ($g$), fundamental ($f$), pressure ($p$), and spacetime ($w$) modes. The 
 $g$, $f$ and $p$ modes are related to the fluid oscillations while $w$ modes correspond to the oscillations of spacetime itself~\cite{Kokkotas:1992xak}. More explicitly speaking, $f$ mode and $p$ mode result from pressure forces, with $f$ mode a particular zero-radial-node mode of $p$ mode in its oscillation eigenfunction. Instead, the $g$-mode is originated from buoyancy forces and thus is excited if the star has temperature/composition gradients or density discontinuities~\cite{finn1986, finn1987,finn1988,1992ApJ...395..240R}. Thus, in the studies of cold compact stars where $g$ modes are absent for conventional neutron stars and quark stars, GW observations of $g$ mode have been studied for hybrid stars as a discriminating signature~\cite{Flores:2013yqa, Kumar:2021hzo, Zhao:2022tcw, Zhao:2022toc}. Since CrSs also have density discontinuities, we expect $g$ mode also be an observational signature of cold CrSs. Besides, $g$ mode is closely related to the Brunt-Väisälä frequency, which is proportional to the adiabatic index deviations $(1/\Gamma_0-1/\Gamma)$ with $\Gamma_0$ the unperturbed value. Thus, $g$ mode only can be expected at the interface of density discontinuities where $\Gamma\neq \Gamma_0$. However, for rapid conversions, a displaced fluid element adjusts its composition to its surroundings almost instantaneously so that $\Gamma$ must equal its unperturbed value $\Gamma_0$.  Therefore, the $g$-mode frequency vanishes for the rapid-conversion scenario as in the conventional hybrid star case~\cite{Tonetto:2020bie}. Thus, for the $g$-mode study, we are more interested in the slow-conversion scenario.

In this asteroseismology study of (cold, non-rotating) CrSs, we first explore the fundamental mode of radial oscillations for the interest of stellar stability, and then the $g$, $f$, the first $p$ modes ($p_1$) of non-radial oscillations for the interest of possible GWs probe, where we adopt the Cowling approximation~\cite{Cowling:1941} that the fluid would oscillate on a fixed background metric with the metric perturbation ($w$ mode) being neglected. For this reason, there is no damping, so the eigenfrequencies of oscillation modes only have the real parts. Note that compared to full treatments using linearized equations of general relativity, typically Cowling approximations have differences by less than $20\%$ for $f$ modes, around $10\%$ for $p$-modes~\cite{Yoshida:1997bf}, and $5\%$--$10\%$ for $g$-modes~\cite{Grig,Tonetto:2020bie,Zhao:2022toc}.

\section{A Recap on the EOS and $M$-$R$ Relation}
Interacting QM includes effects from strong interactions that can modify its behavior, such as one-gluon-exchange induced QCD corrections or color superconductivity. Following Ref.~\cite{Zhang:2020jmb},  the general QM EOSs can be parametrized into the form %~\cite{Zhang:2020jmb}
\be
P=\frac{1}{3}(\rho-4B)+ \frac{4\lambda^2}{9\pi^2}\left(-1+\text{sgn}(\lambda) \sqrt{1+3\pi^2 \frac{(\rho-B)}{\lambda^2}}\right),
\label{eos_tot}
\ee 
where $\lambda$ is related to the size of interquark effects and $B$ is the effective bag constant that accounts for the QCD vacuum contribution. For simplicity, we assume no superconductivity effect in this work. Then we have $\lambda=0$ for $ud$QM, given the negligibly small mass of $u$, $d$ quarks. The EOS of $ud$QM ($\lambda=0$) thus reduces to $P=1/3(\rho-4B)$. For SQM, $\lambda=-\sqrt{3}m_s^2/(4\sqrt{ a_4})$ ~\cite{Zhang:2020jmb, Zhang:2021fla, Zhang:2021iah} with the strange quark mass $m_s=95 \,\rm MeV$~\cite{ParticleDataGroup:2014cgo}.  Here, $a_4$ is to account for the QCD corrections, where $a_4=1$ corresponds to no QCD corrections and smaller values of $a_4$ account for higher-order contributions~\cite{Fraga:2001id,Alford:2004pf,Weissenborn:2011qu}.   
%The energy per baryon number and chemical potential for the interacting QM are~\cite{Zhang:2020jmb}
%\begin{eqnarray}
%\varepsilon_{Q}&=&\frac{3\sqrt{2} \pi}{(\xi_4 a_4)^{1/4}}\frac{ {B}^{1/4}}{\sqrt{(\lambda^2/B+\pi^2)^{1/2}+\lambda/ \sqrt{B}}}\,, \label{EA}
%\end{eqnarray}
The properties of QM are then fully determined by the two parameters $(B, a_4)$. The stability conditions at zero pressure set constraints on the allowed parameter space. Referring to Ref.~\cite{Zhang:2022pse}, we choose a typical benchmark set $B=(20, 35, 50) \rm \, MeV/fm^3\approx ( 111^4, 128^4, 140^4)  \rm \, MeV^4$ for both $ud$QM and SQM.  The allowed range of $a_4$ is given by $a_{4, min}^{ud\text{QM}}\approx(0.35, 0.62, 0.88)$ and $a_{4, max}^{ud\text{QM}}\approx (0.40, 0.70, 1.0)$ for $ud$QM; $a_{4, min}^{\text{SQM}} \approx(0.32, 0.49, 0.64)$ and $a_{4, max}^{\text{SQM}} \approx(0.35, 0.62, 0.88)$ for SQM. 

The phase transition from QM to HM inside CrSs is determined by the crossing point of chemical potentials for the two matter phases~\cite{Annala:2017tqz} at transition pressure $P_{\rm trans}$. 
The chemical potential of interacting QM can be derived as~\cite{Zhang:2020jmb}
\be
\mu_{\rm Q}=\frac{3\sqrt{2}}{(a_4 \xi_4)^{1/4}}\sqrt{[(P+B)\pi^2+\lambda^2]^{1/2}-\lambda}\,,
\label{muP_analy}
\ee
where $\xi_4= \big[ \left(\frac{1}{3}\right)^{\frac{4}{3}}+ \left(\frac{2}{3}\right)^{\frac{4}{3}}  \big]^{-3}\approx1.86$ for $ud$QM and $\xi_4=3$ for SQM. 
For the hadronic sector, we choose Akmal, Pandharipande, and Ravenhall (APR) model~\cite{{Akmal:1998cf,aCompose}} as a benchmark example, considering the fact that CrSs with various hadronic EOSs share similar properties as Ref.~\cite{Zhang:2022pse} has shown. As shown in the appendix, replacing APR with more realistic EOSs~\cite{bComposeAll,sym13091613,PhysRevC.81.015803,Dexheimer:2008ax,dexheimer_2017} 
that including $\varDelta$ resonances and hyperons led to results with similar essential features. Quantitatively, they affect the results at $O(10\%)$ level with main effects happen at the transition points (due to different $P_{\rm trans}$) and maximum mass points (due to the large proportion of the large hadronic core).

%\tred{To add: CSS parameter set}

\begin{figure*}[htb]
\centering
\includegraphics[width=8cm]{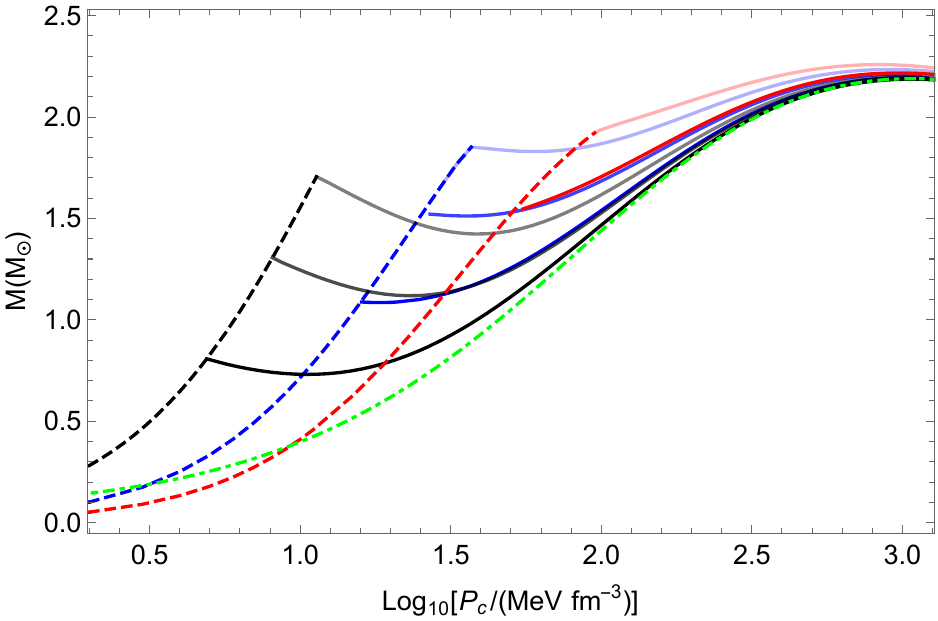}   
\includegraphics[width=8cm]{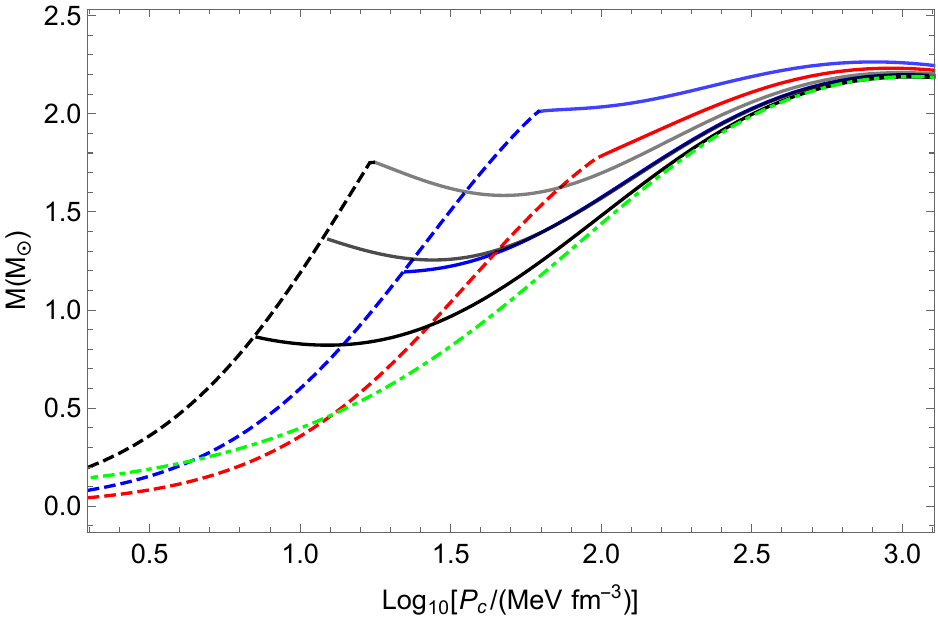}   
 \caption{Masses versus center pressure $P_{\rm c}$ for CrSs of APR HM EOS with $ud$QM (left) and SQM (right) of $B=20$ (black), 35\,(blue), 50\,(red)\,$\rm MeV/fm^3$, and $a_4=a_{4,\rm min}$, $(a_{4,\rm min}+a_{4,\rm max})/2$, $a_{4,\rm max}$ from the darker to lighter color.  APR EOS is applied for the hadronic composition, with the green dot-dashed curve denoting NSs with APR EOS. Dashed lines denote pure QSs.  }  
 \label{fig_pM}
\end{figure*}

The static spherically symmetric background spacetime has the following line element
\begin{equation}
ds^{2}=-e^{ 2\Phi(r)} dt^{2} + e^{ 2\Lambda(r)} dr^{2} + r^{2}(d\theta^{2}+\sin^{2}{\theta}d\phi^{2}),
\end{equation}
where $
  e^{-2\Lambda(r)} = 1 - \frac{2m(r)}{r} \,. $
To obtain the configuration of CrSs, we incorporate the combined EOS of the two matter phases into the Tolman-Oppenheimer-Volkov (TOV) equation~\cite{Oppenheimer:1939ne,Tolman:1939jz}
 \bea
 \begin{aligned}
{dP(r)\over dr}&=-{\left[m(r)+4\pi r^3P(r)\right]\left[\rho(r)+P(r)\right]\over r(r-2m(r))}\,,\,\,\\
{dm(r)\over dr}&=4\pi\rho(r)r^2\,, \\
\frac{d\Phi}{dr} &= - \frac{1}{\rho+ P} \frac{dP}{dr}, 
\end{aligned}
\label{eq:tov}
\eea
with  the boundary conditions
\begin{equation}
  \label{eq:r0}
 \rho(0) = \rho_{\mathrm{c}}\,,\qquad  \Phi(R) = -\Lambda(R)\,,
  \end{equation}
where the star's radius $R$ and mass $M$ are determined by the conditions $P(R)=0$ and $m(R)=M$, respectively.
In Fig.~\ref{fig_pM}, we reproduce the TOV solutions of Ref.~\cite{Zhang:2022pse} in terms of $M$-$P_c$ for the benchmark cases, most of which can meet various astrophysical constraints such as maximum masses $M_{\rm TOV}\gtrsim 2 M_{\odot}$~\cite{Antoniadis:2013pzd,NANOGrav:2019jur,Romani:2022jhd}, tidal deformabilities $\Lambda_{1.4 M_{\rm \odot}}<800$~\cite{LIGOScientific:2017vwq}, as well as NICER mass and radius constraints~\cite{Riley:2019yda,Miller:2019cac,Riley:2021pdl,Miller:2021qha}. At low center pressure, the solutions are those of pure quark stars (dashed lines). As the center pressure increases beyond transition point $P_{\rm trans}$ where the chemical potentials cross, a hadronic core develops, and the solutions yield the configuration of CrSs (solid lines). Note that similar to conventional hybrid stars~\cite{Alford:2013aca}, twin-star (stars with the same mass but different radii) configurations appear for cases with large density discontinuity over transition density ratio $\Delta \rho/\rho_{\rm trans}$.

Besides, as Ref.~\cite{Zhang:2022pse} noted, less parameter space exists to realize stable CrSs in the SQM hypothesis in general due to a larger $P_{\rm trans}$ for SQM compared to $ud$QM, and thus we see fewer solid lines on the right column of Fig.~\ref{fig_pM} than the left.

\section{Radial Oscillations}
To investigate the radial stability, we assume that a fluid element is displaced from its equilibrium position $r$ to $r+\Delta r$, and that
such a perturbation has harmonic time dependence $e^{i\omega t}$.  
The equations for solving infinitesimal radial oscillations $\xi=\Delta r/r$ and the corresponding pressure perturbations $\Delta P$ of a spherical object are~\cite{chanmugam1977radial, vath1992radial, Pereira:2017rmp}
\begin{eqnarray}
\frac{d\xi}{dr}&=&\V(r)\xi+\W(r)\Delta P, \label{eqXi} \\
\frac{d\Delta P}{dr}&=& X(r) \xi + Y(r)  \Delta P, \label{eqDP}
\end{eqnarray}
with the coefficients given by
%
%\begin{eqnarray}
\be
\begin{aligned}
\V(r) &= -\frac{3}{r}-\frac{dP}{dr}\frac{1}{(P+\rho)},  \\
\W(r) &= -\frac{1}{r}\frac{1}{\Gamma P},  \\
X(r) &= \omega^{2}e^{2\Lambda-2\Phi}(P+\rho)r-4\frac{dP}{dr} \\
 &  + \bigg(\frac{dP}{dr}\bigg)^{2}\frac{r}{(P+\rho)}-8\pi e^{2\Lambda}(P+\rho) Pr , \\
Y(r) &= \frac{dP}{dr}\frac{1}{(P+\rho)}- 4\pi(P+\rho)r e^{2\Lambda}, \label{ecuacionparaP}
\end{aligned}
\ee
%\end{eqnarray}
where $ \Gamma=\frac{n}{P}\frac{dP}{dn}=\frac{\rho+P}{P}\frac{dP}{d\rho}$ is the adiabatic index. The initial conditions are $
(\Delta P)_{r=0}=-3(\xi \Gamma P)_{r=0}$ and $\xi(0)=1$. We obtain the eigenfrequencies by solving Eq.~(\ref{eqXi}) and Eq.~(\ref{eqDP}) using the shooting method, such that the boundary condition
\begin{equation}\label{PenSuperficie}
(\Delta P)_{r=R}=0
\end{equation}
is satisfied. 

For the scenario of slow conversions, the matching condition across the interface of the two matter phases is
\begin{equation}
[\xi]^+_- = 0 , \quad [\Delta P]^+_-=0,
\label{junction_xi_slow}
\end{equation}
where $[x]^+_-\equiv x^+-x^-$.

\begin{figure*}[htb]
\centering
\includegraphics[width=8cm]{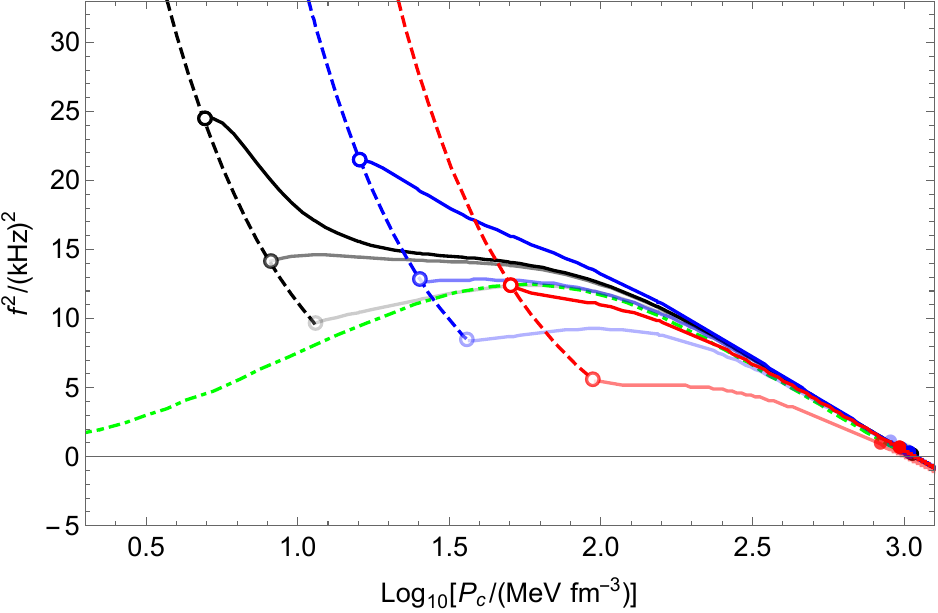}   
\includegraphics[width=8cm]{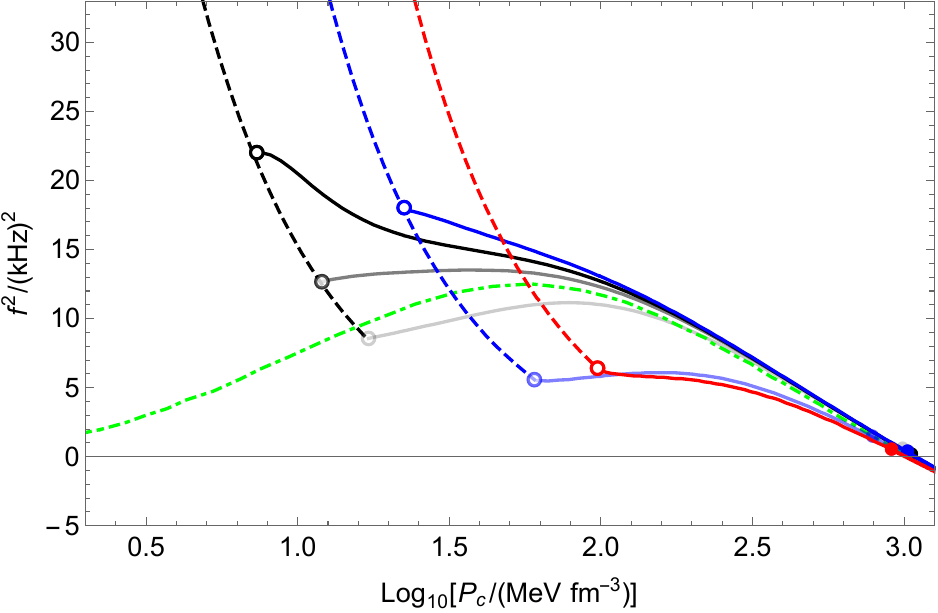}   
\includegraphics[width=8cm]{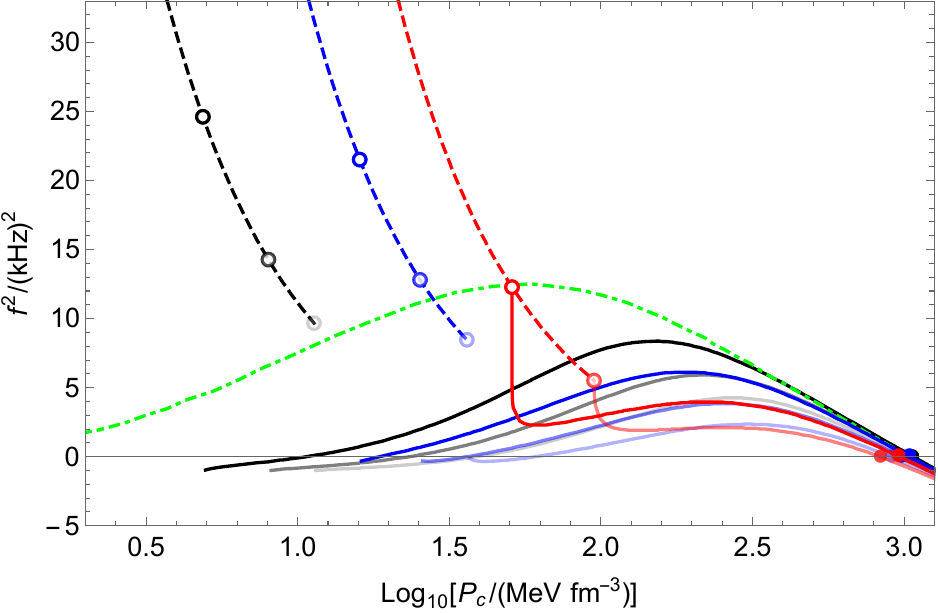}   
\includegraphics[width=8cm]{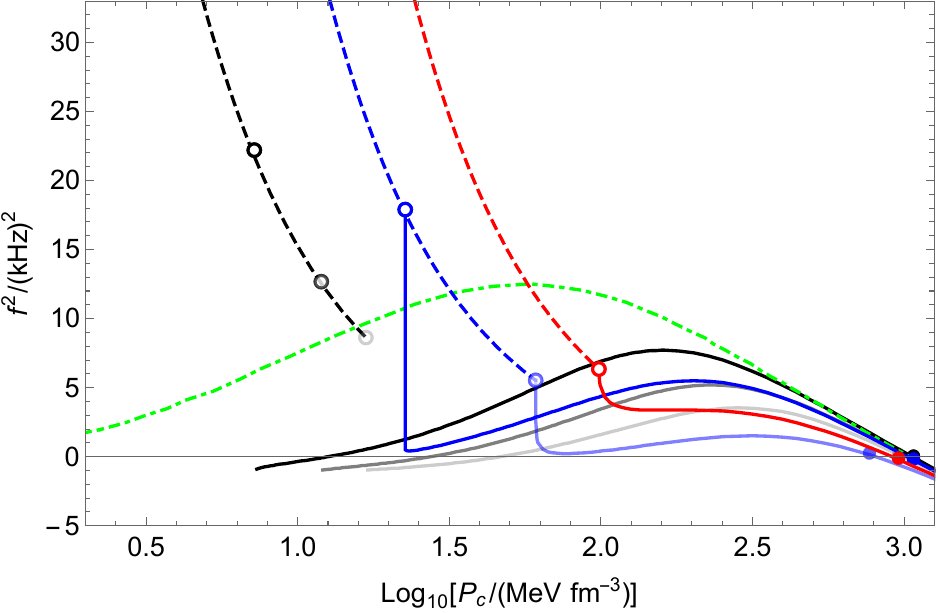}   
 \caption{Radial-oscillation frequency squares for (top) slow conversion and (bottom) rapid conversion of CrSs with $ud$QM (left) and SQM (right) as a function of center pressure.  The line-color convention follows that of Fig.~\ref{fig_pM}.   In addition, we use hollow circles to mark the transition points and filled one to mark the maximum mass points. Those with nearly-vertical lines are non-twin branches,  where  the  eigenfrequencies drop more continuously after the transition point. Those without nearly-vertical lines (i.e. twin branches) signal more abrupt change in frequencies at the transition point.}
   
 \label{fig_radial}
\end{figure*}

For the scenario of rapid conversions, the matching condition at the interface changes to 
\begin{equation}
[\Delta P]^+_-=0      , \quad \left[\xi -\frac{\Delta P}{r P'} \right]^+_-=0,
\label{junction_xi_rapid}
\end{equation}
where primes on the variables denote the partial derivatives with respect to $r$.
We show the results in Fig.~\ref{fig_radial} for both slow conversion and rapid conversion scenarios, where one can see quantitative and qualitative differences regarding the behavior of the fundamental-mode frequencies. In general, the frequencies for slow conversions are noticeably larger than those of rapid conversions, as expected from the conventional hybrid star studies~\cite{Pereira:2017rmp,Goncalves:2022phg}. 

We see that for the slow conversion scenario, similar to conventional hybrid stars, all lines of $f^2(P)$ are continuously connected and positive before the maximum mass points, even for regions with negative $\partial M/\partial P_c$. This trend leads to the possibility of triplet stars (three stellar configurations that have the same mass but different radii) at the intermediate center pressure region. In contrast to hybrid star cases where the triplet stars are one neutron star and two hybrid stars, in the case here, the triplets are one quark star and two CrSs. Interestingly, the radial oscillation frequencies of CrSs are always larger than those of QSs at any center pressure, approaching those of NSs at high center pressure, thus showing an inverted trend compared to conventional hybrid stars~\cite{Goncalves:2022phg}. 

For the case of rapid conversions, we observe an abrupt drop of oscillation frequency square $f^2$ to negative values (i.e., unstable regime) at $P_{\rm trans}$ where $\partial M/ \partial P_c$ turns negative (referring to Fig.~\ref{fig_pM} for twin-star branches), then slowly increases to positive values (i.e., stable regime) where $\partial M/ \partial P_c$ turns back positive. Similar features also have been found for conventional hybrid stars~\cite{Pereira:2017rmp,Lugones:2021zsg,Goncalves:2022phg}.  However, compared to conventional hybrid stars, the low-mass (i.e., small $P_c$) branches are pure quark stars rather than pure NSs for CrSs studies, thus featuring much larger frequencies.

For both conversion scenarios (slow and rapid), due to the larger $P_{\rm trans}$ determined by the chemical potential crossing, CrSs of $ud$QM and SQM with larger $a_4$ for a given bag constant has lower oscillation frequencies. Due to the same reason, CrSs with SQM have lower oscillation frequencies than CrSs with $ud$QM of the same bag constant.

\section{Non-radial oscillations}

\begin{figure*}[htb]
\centering
\includegraphics[width=8.5cm]{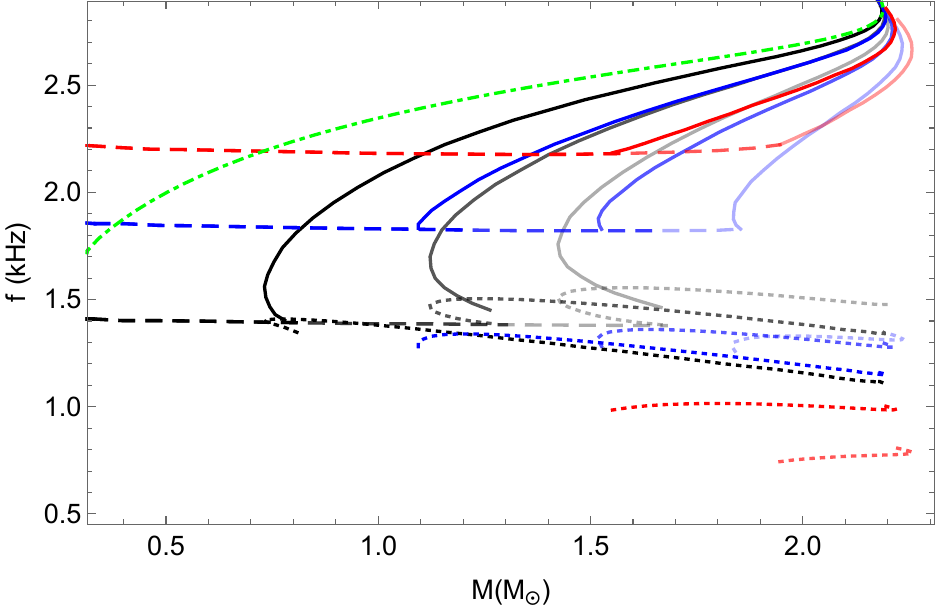}   
\includegraphics[width=8.5cm]{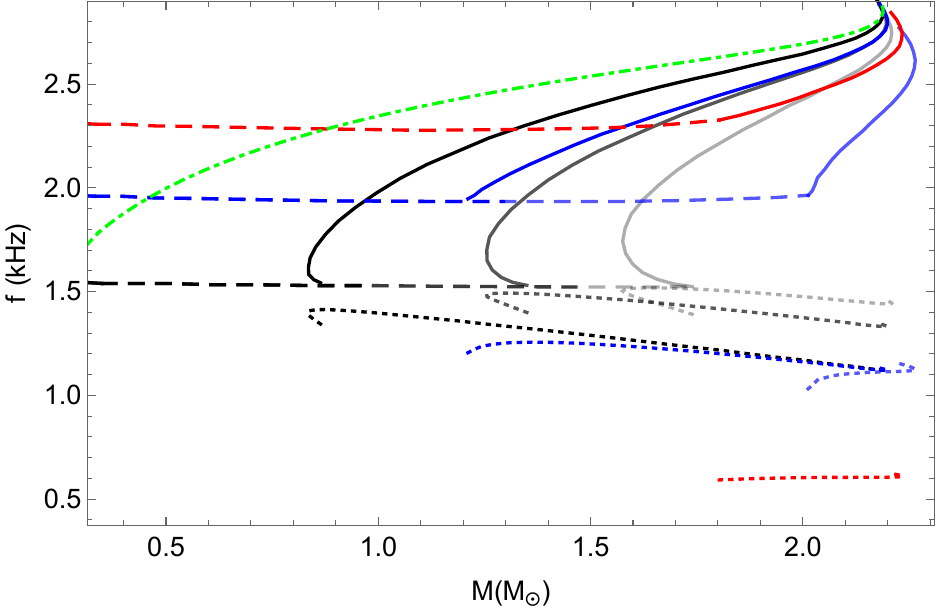}   
\includegraphics[width=8.5cm]{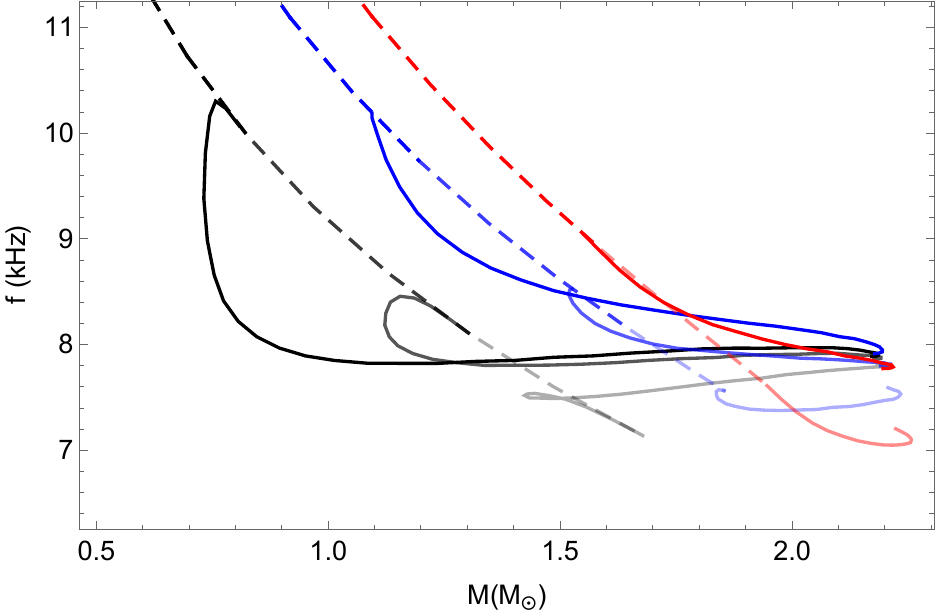}   
\includegraphics[width=8.5cm]{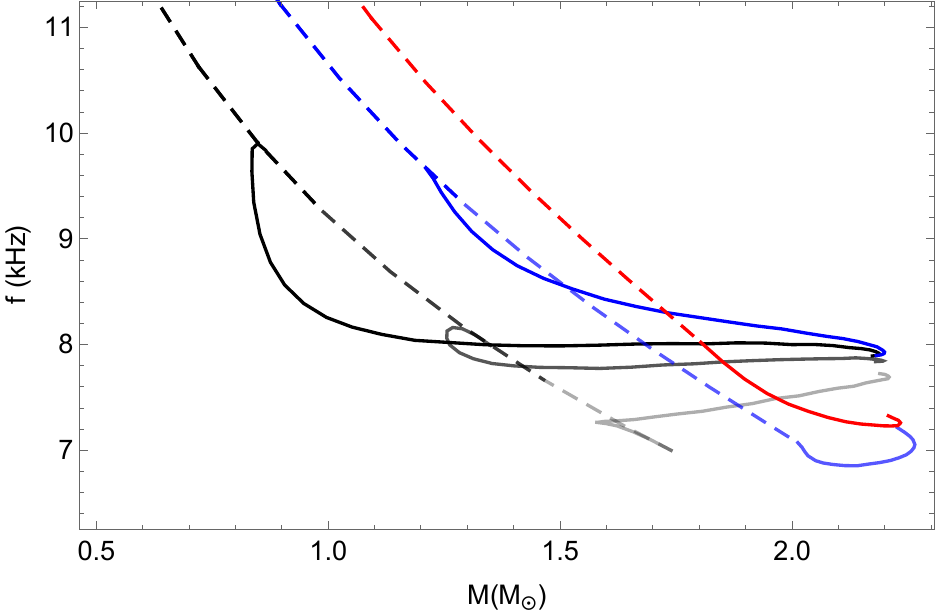}   
 \caption{Non-radial oscillation frequencies of (top) $f$ modes (solid lines) and $g$ modes (dotted lines), and (bottom)  $p_1$ modes (solid lines) for CrSs with $ud$QM (left) and SQM (right) as a function of star mass. The line-color convention follows that of Fig.~\ref{fig_pM}. }  
 \label{fig_nonradial}
\end{figure*}

For non-radial oscillations of compact stars, the fluid Lagrangian displacement vector is
given by~\cite{Sotani:2010mx}
%-----------------------------------
\begin{equation}
  \label{eq:nonradialGR-1}
  \xi^i = \left(e^{-\Lambda}W, -V\partial_\theta,
  -V\sin^{-2}\theta\partial_\phi\right)r^{-2}Y_{\ell m} \,,
\end{equation}
%-----------------------------------
where $W$ and $V$ are functions of $t$ and $r$, while $Y_{\ell m}$ is the
spherical harmonic function.  Then the perturbation of the four-velocity,
$\delta u^\mu$, can be written as
%-----------------------------------
\begin{equation}\label{eq:nonradialGR-2}
  \delta u^\mu = \left(0, e^{-\Lambda}\partial_t W, -\partial_t V
  \partial_\theta, -\partial_t
  V\sin^{-2}\theta\partial_\phi\right)r^{-2}e^{-\Phi}Y_{\ell m}\,.
\end{equation}
%-----------------------------------
With these variables, the perturbation equations describing the fluid oscillations can be obtained by taking a variation of the energy-momentum conservation law.
Assuming a harmonic dependence on time for the perturbative variables $W(t,r)=W(r)e^{i\omega t}$ and $V(t,r) = V(r)e^{i\omega t}$,  the fluid perturbations are determined from the following equations~\cite{Sotani:2010mx}
\begin{align}
  \frac{dW}{dr} &= \frac{d\rho}{dP}\left[\omega^2 r^2e^{\Lambda-2\Phi}V + \frac{d\Phi}{dr} W\right] - \ell(\ell+1)e^{\Lambda}V, \label{eq1} \\
  \frac{dV}{dr} &= 2\frac{d\Phi}{dr}V-e^\Lambda\frac{W}{r^2},  \label{eq2}
\end{align}
together with Eq.~(\ref{eq:tov}). The boundary condition in the vicinity of stellar center is $W(r)=Cr^{\ell+1}+{\cal O}(r^{\ell+3})$ and $V(r) = -Cr^\ell/\ell+{\cal O}(r^{\ell+2})$, where $C$ is an arbitrary constant.  The eigenfrequencies for each mode following the order $\omega_g<\omega_f<\omega_{p1}$ are obtained by solving  Eq.~(\ref{eq1}) and Eq.~(\ref{eq2})  using the shooting method starting from lowest possible trial frequencies, such that the boundary condition
\begin{equation} 
  \omega^2 r^2e^{\Lambda-2\Phi}V + \Phi' W = 0
\end{equation}
is satisfied at star surface $r=R$.
The additional junction conditions at the surface of discontinuity in the star interiors are
\be
\begin{aligned}
  W_+ &= W_-, \\
  V_+ &= \frac{e^{2\Phi}}{\omega^2 {r_d}^2}\big[\frac{\rho_-+P}{\rho_++P}
       \left(\omega^2{r_d}^2e^{-2\Phi}V_-+e^{-\Lambda}\Phi' W_-\right) \\
      &-e^{-\Lambda}\Phi' W_+\big],
\end{aligned}
\ee
where $r_d$ denotes the radial position of the density discontinuity $\Delta \rho$. Variables with subscript of minus and plus signs denote their values at $r=r_d-0$ and $r=r_d+0$, respectively. We obtain the results of quadrupole oscillations ($\ell=2$) shown in Fig.~\ref{fig_nonradial}.  

%As we have shown in Fig.~\ref{fig_radial}, all configurations below the maximum masses are stable, with an additional small portion of stable region even extended beyond maximum masses for slow conversion scenario.

For $f$ and the $p_1$ modes, we see their eigenfrequencies sit in between those of neutron stars and pure quark stars in $1.4$--$2.8$ kHz and $7$--$10$ kHz, respectively. As the center pressure passes the transition pressure and the resulting CrS mass increases, the eigenfrequencies deviate from those of quark stars, approaching those of neutron stars as the star masses reach their maxima. We see that it is hard to infer the type of compact stars using $f$ modes, where different types have overlapping frequencies in a narrow frequency range. However, since the $p_1$-mode frequencies of quark stars are significantly above those of neutron stars at low and intermediate-mass ranges, this large difference translates to larger frequencies ($\sim 8-10$ kHz) and an inverting trend as mass increases for major parts of the $p_1$-mode GWs of CrSs, compared to the $p_1$ modes of conventional hybrid stars ($\sim 4$--$7$ kHz~\cite{Miniutti:2002bh,Flores:2013yqa,Thapa:2023grg} for hybrid stars with $M_{\rm TOV}\gtrsim 2 M_{\rm \odot}$) where the frequencies always grow for increasing masses. These results provide a possible way to discriminate between CrSs and conventional hybrid stars. As shown in the appendix, the variations of hadronic EOSs (e.g. adding hyperons and $\Delta$ baryons) only affect our $p$-mode results at $O(10\%)$ level with maximal changes around maximum mass points, considering $p$-mode oscillation dominates in the regions close to the star surface~\cite{Andersson:1997rn,Sotani:2010mx,Kunjipurayil:2022zah}, which are composed of mostly QM for CrSs except for very large mass configurations where the large hadronic core is not far from the star surface.

From Fig.~\ref{fig_nonradial}, we see that the $g$-mode frequencies of CrSs with large bag constant (red dotted lines) are $\lesssim 1 \rm \, kHz$, within the sensitive range of current GW detectors, thus serve as an appropriate observable. Note that CrSs with $B=50 \rm \, MeV/fm^3$ have lowest $g$-mode frequencies around $0.5 \rm \, kHz$, which are at similar range as those of conventional hybrid stars~\footnote{Higher $g$-mode frequencies are possible if one considers slow stable hybrid stars (SSHSs) where stability window extends well beyond maximum mass point due to possible slow conversion~\cite{Tonetto:2020bie}. For the very extended branches, the transition point becomes the maximum mass point with stability extending to $10\sim50$ times nuclear saturation densities, and are shown to break universal relations badly~\cite{Ranea-Sandoval:2023ixr}.}.

Therefore, together with the $p_1$-mode analysis above, we see a direct way to discriminate CrSs from other types of compact stars:  compact stars emitting GWs within $0.5$--$1$ kHz\footnote{This is a subset of $g$-mode frequencies shown in Fig.~\ref{fig_nonradial}, in order to be more conservative and avoid the overlapping region with $f$ modes.}  should be interpreted as CrSs or conventional hybrid stars rather than neutron stars or pure quark stars, and a further GW detection of $p_1$ modes above 8 kHz or identification of a decreasing trend of frequencies versus star masses will help identify it to be CrSs rather than conventional hybrid stars.  
 
\begin{figure}[htb]
\centering
\includegraphics[width=8cm]{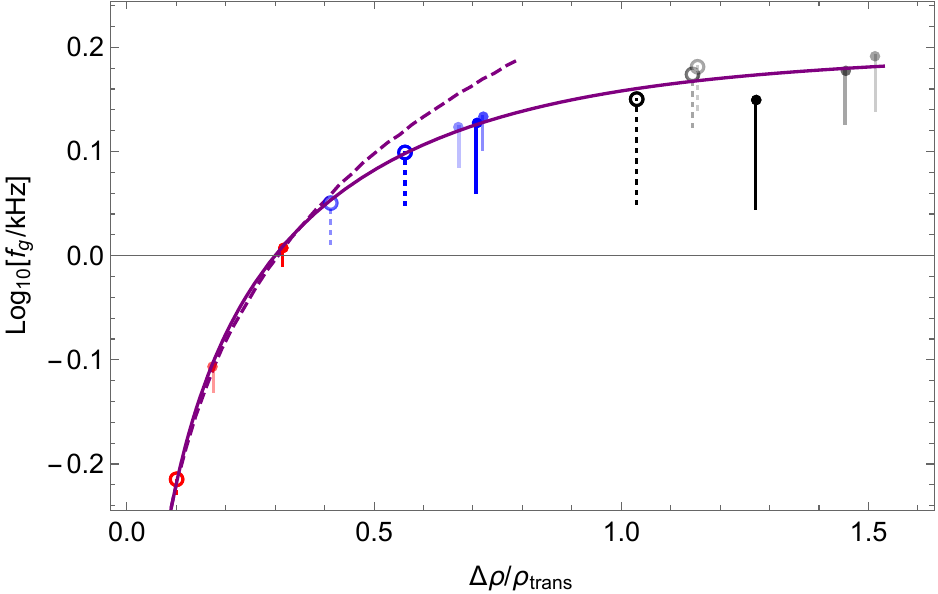}   
 \caption{${\Delta\rho}/{\rho_{\rm trans}}$ vs $g$-mode frequencies for different ($B$,$a_4$) sets. 
Filled and empty circles denote the maximum $g$-mode frequencies of CrSs with $ud$QM and SQM, respectively, with the color convention being same as Fig.~\ref{fig_pM}, and the vertical bars denote the frequencies ranges. The dashed purple line represents the fit for conventional hybrid stars adapted from Ref.~\cite{Ranea-Sandoval:2018bgu}, while the solid purple line denotes our fit for CrSs.}  
 %Note that we dropped branches with no (stable) CrSs configurations (i.e. no crossing of $\mu_H$ and $\mu_Q$), such as the $(B,a_4)=(50 \rm\, MeV/fm^3, a_4^{\rm max})$ for $ud$QM and $B=50 \rm\, MeV/fm^3$ for SQM. }
 \label{gmode_drho}
\end{figure}

Last but not least, we make some more comments on the behaviour of $f$ mode and $g$ modes of CrSs.  We see that a larger bag constant {$B$}, or equivalently a smaller $\Delta \rho/\rho_{\rm trans}$, maps to a larger $f$-mode frequency, with the opposite trend for $g$ modes. This feature is consistent with the expectation that $f$ mode is positively correlated with average density insides stars~\cite{Benhar:2004xg,Andersson:1997rn,Chirenti:2015dda,Lau:2009bu}, and the finding of Ref.~\cite{Miniutti:2002bh, Ranea-Sandoval:2018bgu} for conventional hybrid stars regarding the positive correlation between density discontinuities and the $g$-mode frequencies.  We show such correlation explicitly in Fig.~\ref{gmode_drho}, with values of  $\Delta \rho/\rho_{\rm trans}$ for each benchmark set listed in Table.~\ref{tab:drho}. Due to similar reasoning, from Fig.~\ref{fig_nonradial}, we see that CrSs with $ud$QM have slightly lower $f$-mode and higher $g$-mode frequencies than those of CrSs with SQM of the same bag constant. 

\begin{table}[]
\leavevmode
\begin{center}  
\resizebox{\columnwidth}{!}{
\begin{tabular}{c|cc|cc|cc}
\hline\hline
\multicolumn{1}{c|}{\multirow{2}{*}{\begin{tabular}[c]{@{}c@{}}$\frac{\Delta\rho}{\rho_{\rm trans}}$\end{tabular}}} & \multicolumn{2}{c|}{$a_{4,min}$} & \multicolumn{2}{c|}{$\frac{a_{4,\rm min}+a_{4,\rm min}}{2}$} & \multicolumn{2}{c}{$a_{4,max}$} \\ \cline{2-7} 
\multicolumn{1}{c|}{}                                                                   & \multicolumn{1}{c|}{$ud$QM}  & SQM & \multicolumn{1}{c|}{$ud$QM}  & SQM & \multicolumn{1}{c|}{$ud$QM} & SQM \\ \hline
$B_{20}$                                                                                     & \multicolumn{1}{c|}{1.27}      &  1.03   & \multicolumn{1}{c|}{1.45}      &  1.14   & \multicolumn{1}{c|}{1.51}     &  1.15   \\
$B_{35}$                                                                                     & \multicolumn{1}{c|}{0.71}      &   0.56  & \multicolumn{1}{c|}{0.72}      &   0.41  & \multicolumn{1}{c|}{0.67}     &  NA   \\
$B_{50} $                                                                                    & \multicolumn{1}{c|}{0.32}      &   0.10  & \multicolumn{1}{c|}{0.17}      &  NA   & \multicolumn{1}{c|}{NA}     &  NA   \\ \hline\hline
\end{tabular}
}
\caption{${\Delta\rho}/{\rho_{\rm trans}}$ for different ($B$,$a_4$) set.   $B_x$ denotes $B\equiv x  \rm \, MeV/fm^3$. The left element in each table entry represents values for CrSs with $ud$QM, while the right one for CrSs with SQM. ``NA" represents ``Not applicable", meaning no CrSs configuration exists for the corresponding ($B$, $a_4$) set. }
\label{tab:drho}
\end{center}  
\end{table}

\begin{figure}[h!]
\centering
\includegraphics[width=8cm]{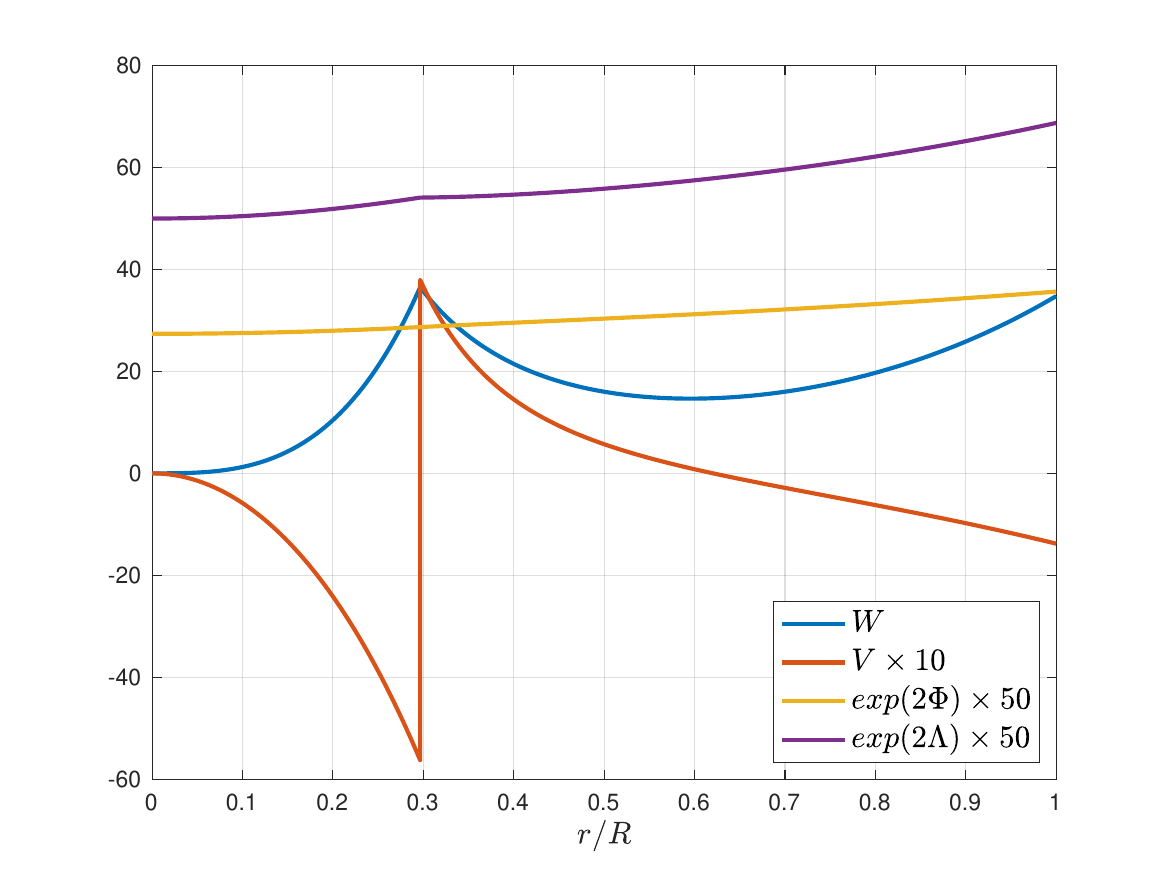}
\includegraphics[width=8cm]{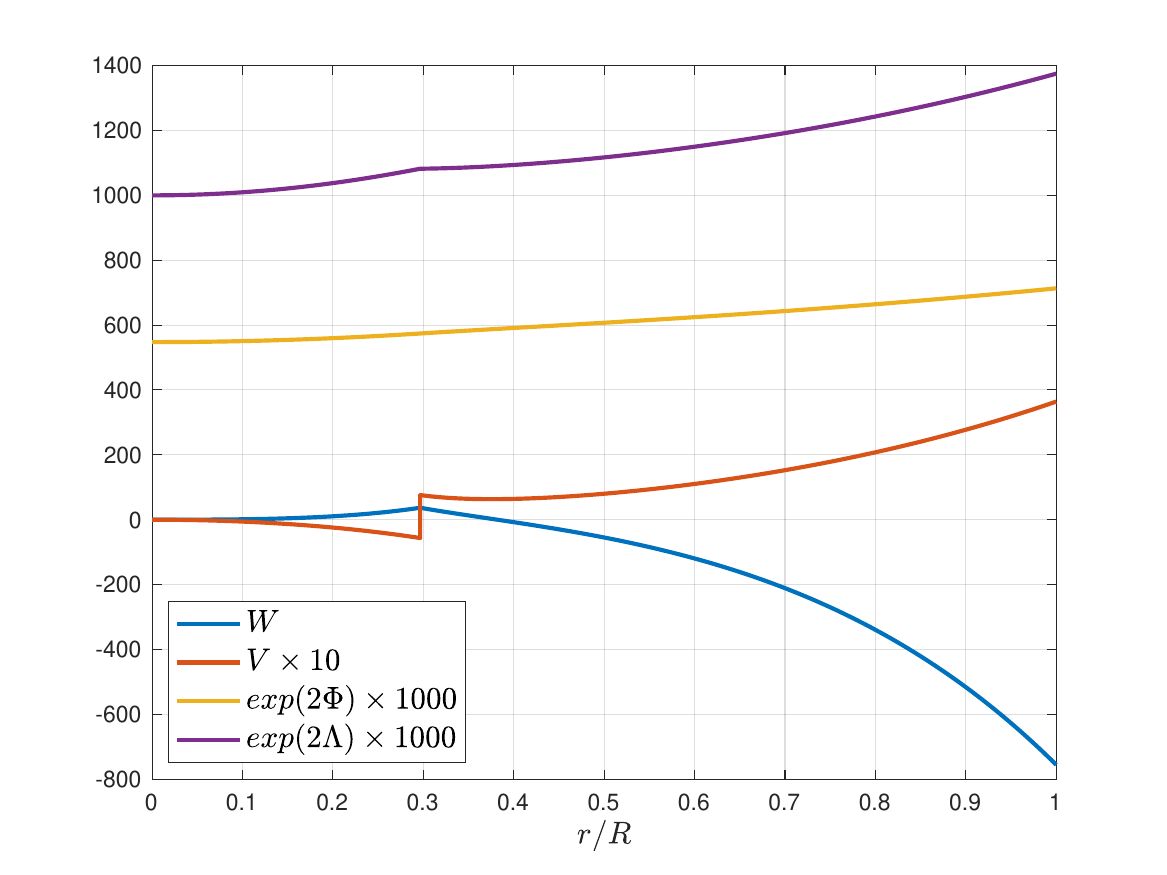}
 \caption{Eigenfunctions of $f$-mode (top) and $g$-mode (bottom) for the twin branch with a frequency jump at the transition point. The CrS has center pressure $15.68 \rm \ MeV\ fm^{-3}$ (star mass $1.60\rm \ M_\odot$), above and close to the transition pressure, consisting of a small hadronic core. The set of parameters is chosen as the lightest black lines in the left top panel of Fig. 3 in the manuscript ($B=20\rm\, MeV/fm^3$, $a_4=a_{4,\rm max}$).}
\label{twin_eigen}
\end{figure}

For very large $\Delta \rho/\rho_{\rm trans}$ (light-colored black solid lines on the top left of Fig.~\ref{fig_nonradial}),  $f$ modes become discontinuous at the transition point with a gap of $\sim0.1$ kHz, which, as we have examined, is a general feature that also manifests for conventional twin hybrid stars with large $\Delta \rho/\rho_{\rm trans}$~\footnote{This discontinuity even presents in full GR treatment~\cite{ZhaoMail}. For a $f$-mode gap at large $\Delta \rho/\rho_{\rm trans}$ in conventional hybrid star setting, see Ref.~\cite{Zhao:2022tcw}. }.  Such a jump in $f$-mode is associated with the abrupt change of behaviours of the eigenfunctions ($W$,$V$), as can be seen in Fig.~\ref{twin_eigen}a. 

Besides, we also see that for these case (those with $f$-mode jumps), the $g$-mode right after the transition point is almost connected to the $f$-mode 
right before the transition point. 
This is also associated the eigenfunction behaviours in that in these cases the $W$, $V$ of $g$-mode behave very similarly to those of $f$-mode of pure stars ($W$ and $V$ are far-separated opposite sides at surface.), as Fig.~\ref{twin_eigen}b shows. Note that we still categorize these as $g$-modes since they are lowest-lying modes and they has the same node structure ($W(r)$ have one node while $V(r)$ has no nodes.) same as those other $g-$modes.

%where each row represents $\Delta \rho/\rho_{\rm trans}$ with $B_x$ denoting $B=x  \rm \, MeV/fm^3$, while different columns represents $\Delta \rho/\rho_{\rm trans}$ with different $a_4$ value. }

\section{Summary}
In this paper, we have studied the radial and non-radial oscillations of Cross stars (CrSs) that have an inverted structure compared to conventional hybrid stars in the context of absolutely stable $ud$QM and SQM hypotheses, respectively. 

Similar to conventional hybrid stars, radial oscillations of CrSs have stability windows in rapid conversions that coincide with the conventional $\partial M/\partial P_c>0$ criteria, yielding disconnected branches for CrSs with large density discontinuities. For slow conversions, these disconnected branches can become connected, thus increasing the stability window of the CrSs configurations and allowing triplet stars in which two CrSs and one pure quark star are of the same mass but very different radii. In general, CrSs with $ud$QM have higher radial oscillation frequencies than those of CrSs with SQM of the same bag constant.

For the non-radial modes, we can utilize $g$-mode pulsations to discriminate CrSs with neutron stars and quark stars in the similar frequencies range $0.5\sim1$ kHz, as those of conventional hybrid stars. A further GW detection on the first $p$ mode above $8$ kHz or identifying decreasing frequencies for increasing masses helps differentiate the compact objects to be CrSs rather than conventional hybrid stars. Besides, in general,  CrSs with $ud$QM have slightly lower $f$-mode, similar $p_1$-mode, and higher $g$-mode frequencies than those of CrSs with SQM of the same bag constant.

 In this study, for simplicity purposes, we utilized Cowling approximations that restrict to stellar oscillations. A full GR treatment with metric perturbations not only gives a more precise description but also can account for additional information, such as the damping of GWs~\cite{Tonetto:2020bie}. Besides, the $f$ mode jumps we found that stars with large $\Delta\rho/\rho_{\rm trans}\, (\gtrsim1)$ are rarely studied even for traditional hybrid stars likely due to the $M_{\rm TOV}\gtrsim 2M_{\odot}$ constraint. However, such a jump indicate the change of behaviours of the eigenfucntions, which is worth to further investigation. Moreover,  thermal effects, magnetic fields, and rapid rotations may also affect the stellar oscillations~\cite{1988ApJ...325..725M,Sotani:2006at,Leung:2022mvm,Kruger:2019zuz,Largani:2021hjo,Dimmelmeier:2005zk,Ng:2020etb,Kruger:2021zta}. On the other hand,  strange quark matter may be localized as solid clusters called strangeons~\cite{Xu:2003xe,Miao:2020cqj,Lai:2022yky}, which can form strangeon stars~\cite{2009MNRAS.398L..31L,Lai:2017ney,Gao:2021uus,Li:2022qql,Zhang:2023mzb,Chen:2023pew}. Our study hints that the hybrid configurations of strangeon stars~\cite{Zhang:2023szb} and their $p$-mode GW asteroseismology may also differ greatly from those of conventional hybrid stars. We leave these for future studies.

\begin{acknowledgments}
\subsection*{Acknowledgments}
We thank Hajime Sotani, Tianqi Zhao, Enping Zhou for valuable discussions. In particular, we greatly appreciate Tianqi Zhao for his help in confirming the correctness of our Cowling results and the consistency with full GR treatments.
 C.~Zhang is supported by the Jockey Club Institute for Advanced Study at The Hong Kong University of Science and Technology. 
Y. L. is supported by the Boya Fellowship of Peking University and the National Natural Science Foundation of China (No. 12335009).
     H.-B. Li and L. Shao are supported by the National SKA Program of China (No. 2020SKA0120300), the National Natural Science Foundation of China (No. 11975027, No. 11991053), and the Max Planck Partner Group Program funded by the Max Planck Society. R.X Xu is supported by the National SKA Program of China (No. 2020SKA0120100). 
     
     Chen Zhang and Yudong Luo contributed equally to this work.
\end{acknowledgments}

\setcounter{equation}{0}
\setcounter{figure}{0}
\setcounter{table}{0}
\setcounter{page}{1}
\makeatletter
\renewcommand{\thefigure}{A\arabic{figure}}
%\renewcommand{\bibnumfmt}[1]{[S#1]}
%\renewcommand{\citenumfont}[1]{S#1}

%\section{Appendix}
%\bigskip

\section{Appendix: Hadronic EOSs with hyperons and $\varDelta$ baryons}
\label{app_hyperDelta}
 In Fig.~\ref{global} and Fig.~\ref{osc} below, we show the explicit results for CrSs with hadronic EOSs, e.g., DD2YDelta1.1~\cite{aDD2Y1.1}, 
 DD2YDelta1.3~\cite{bDD2Y1.3}, DS(CMF)-7~\cite{cDSCMF7}, DS(CMF)-8~\cite{dDSCMF8} that include both hyperons and $\varDelta$ resonances, as expected to exist at the high-density regime of the hadronic matter phase. The top panel of Fig.~\ref{global} shows that these $\varDelta$-included hyperonic examples happen to have very similar $P_{\rm trans}$, as determined by the crossing of the chemical potentials of quark matter phase (dashed black line) and hadronic matter phase (solid colored lines). The bottom panel of Fig.~\ref{global} shows that the resulting CrSs all satisfy recent NICER constraints. These results are not very different from those with APR EOS and the same SQM EOS~\cite{Zhang:2022pse}. 

  For the results of nonradial oscillations, we can see from  Fig.~\ref{osc} that compared to the results using APR and the same SQM EOS (the darkest black lines in right panels of Fig.~\ref{fig_nonradial}), their qualitative behaviours are quite similar, while the exact numerical results have only $O(10\%)$ level differences, with main effects occur at the transition points (due to different $P_{\rm trans}$) and the maximum mass points (due to the large proportion of the large hadronic core). 
  
 %\tred{All these justify our use of APR as the benchmark EOS in this proof-of-concept work.}
\begin{figure}[htb]
\centering
\includegraphics[width=0.5\textwidth]{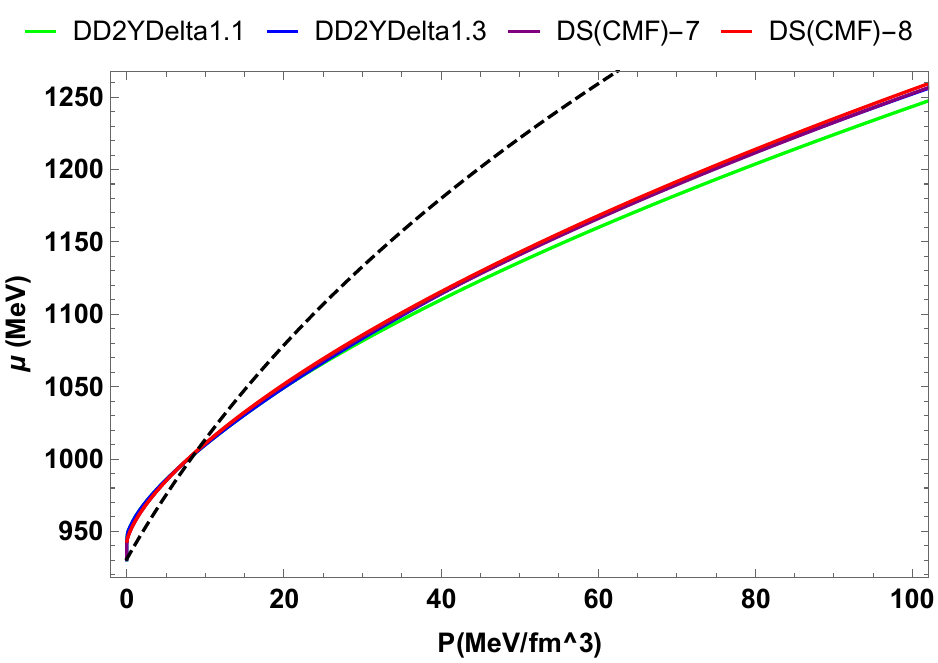}
\includegraphics[width=0.49\textwidth]{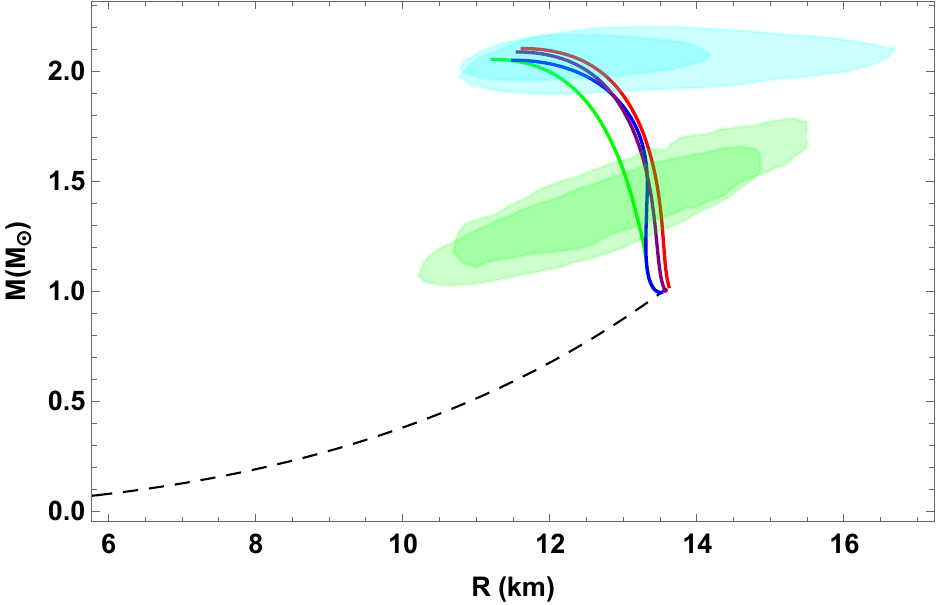}   
 \caption{(Top) Chemical potentials $\mu(P)$ of the hyperonic EOSs, with SQM (black dashed) of $B=20\rm \, MeV/fm^3$, $a_4=a_{4, \rm min}$, and the hadronic EOSs are DD2YDelta1.1 (green)~\cite{aDD2Y1.1}, DD2YDelta1.3 (blue)~\cite{bDD2Y1.3}, DS(CMF)-7 (purple)~\cite{cDSCMF7}, DS(CMF)-8 (red)~\cite{dDSCMF8}, as the top legends shows. (Bottom) The mass-radius curves of the resulting CrSs (solid colored lines), with the same color convention as the top panel. The shaded regions are the constraints with $90\%$ credibility from the NICER mission PSR J0030+0451 (green colored) \cite{NICER1,NICER2}, PSR J0740+6620 (cyan colored)\cite{NICER3,NICER4}.}
\label{global}
\end{figure} 

\begin{figure}[htb]
\centering
\includegraphics[width=0.49\textwidth]{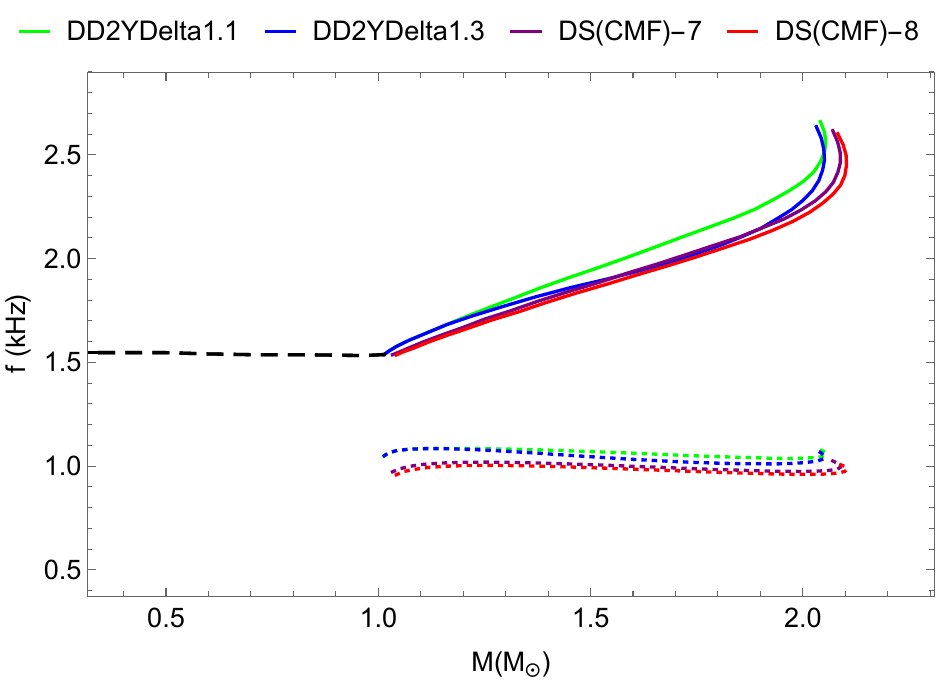}
\includegraphics[width=0.49\textwidth]{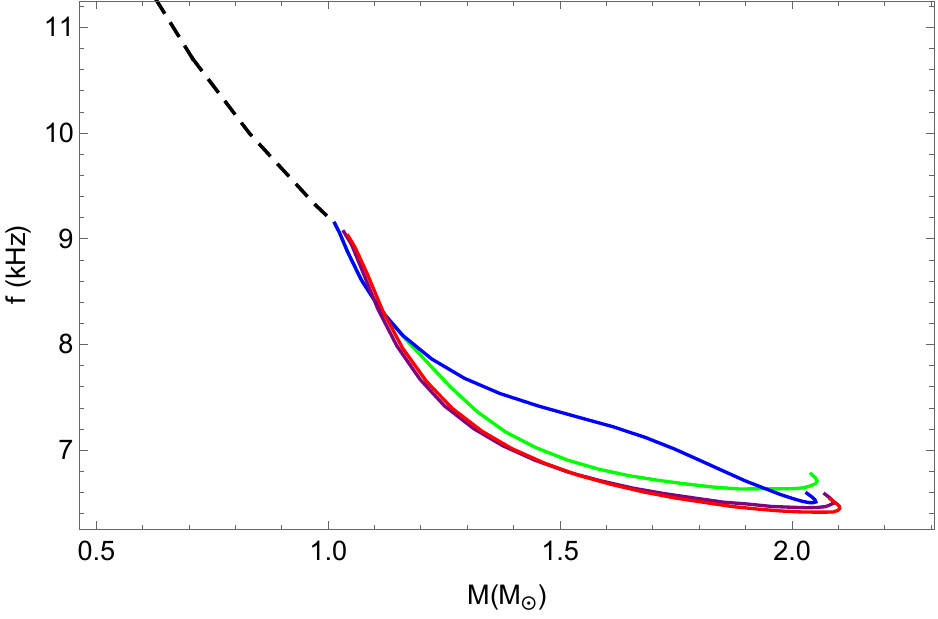}   
 \caption{(Top) Frequencies of $f$ modes (solid lines) and $g$ modes (dotted lines). (Bottom) $p_1$ modes (solid lines) for CrSs with the same parameter choices and color style conventions as Fig.~\ref{global}.}
\label{osc}
\end{figure} 

%SQM of $B=20\rm \, MeV/fm^3$, $a_4=a_{4, \rm min}$ and hadronic EOSs of  DD2YDelta1.1 (green), DD2YDelta1.3 (blue), DS(CMF)-7 (purple), DS(CMF)-8 (red)
\bibliography{OscCrS_Refs}

\end{document}